\documentclass[aps,12pt,showpacs,preprint,groupedaddress,floatfix]{revtex4}
\usepackage{graphicx}
\usepackage{dcolumn}
\usepackage{bm}
\usepackage{amssymb}
\usepackage{amsmath}
\usepackage{color}

\begin{document}

\title{Global analysis of quadrupole shape invariants based on covariant energy density functionals}
\author{S. Quan}
\author{Q. Chen}
\author{Z. P. Li}\email{zpliphy@swu.edu.cn}
\affiliation{School of Physical Science and Technology, Southwest University, Chongqing 400715, China}

\author{T. Nik\v si\'c}
\author{D. Vretenar}
\affiliation{Physics Department, Faculty of Science, University of Zagreb, 10000 Zagreb, Croatia}

\bigskip
\date{\today}

\begin{abstract}

\begin{description}
\item[Background] Coexistence of different geometric shapes at low energies presents a universal structure phenomenon that occurs over the entire chart of nuclides. Studies of the shape coexistence are important for understanding the microscopic origin of collectivity and  modifications of shell structure in exotic nuclei far from stability.

\item[Purpose] The aim of this work is to provide a systematic analysis of characteristic signatures of coexisting nuclear shapes in different mass regions, using a global self-consistent theoretical method based on universal energy density functionals and the quadrupole collective model.

\item[Method] The low-energy excitation spectrum and quadrupole shape invariants of the two lowest $0^{+}$ states of even-even nuclei are obtained as solutions of a five-dimensional collective Hamiltonian (5DCH) model, with parameters determined by constrained self-consistent mean-field calculations based on the relativistic energy density functional PC-PK1, and a finite-range pairing interaction.

\item[Results] The theoretical excitation energies of the states: $2^+_1$, $4^+_1$, $0^+_2$, $2^+_2$, $2^+_3$, as well as the $B(E2; 0^+_1\to 2^+_1)$ values, are in very good agreement with the corresponding experimental values for 621 even-even nuclei. Quadrupole shape invariants have been implemented to investigate shape coexistence, and the distribution of possible shape-coexisting nuclei is consistent with results obtained in recent theoretical studies and available data.

\item[Conclusions] The present analysis has shown that, when based on a universal and consistent microscopic framework of nuclear density functionals, shape invariants provide distinct indicators and reliable predictions for the occurrence of low-energy coexisting shapes. This method is particularly useful for studies of shape coexistence in regions far from stability where few data are available. 
\end{description}
\end{abstract}


\maketitle


Shape coexistence presents an intriguing phenomenon in mesoscopic systems in which low-energy nearly degenerate states occur characterized by different geometrical shapes. Atomic nuclei, in particular, often exhibit (sets of) energy eigenstates with very different electromagnetic properties: moments and transition rates, and different distributions of proton and neutron pairs with respect to their corresponding Fermi levels \cite{Heyde11,Wood92,Heyde83}. The origin of nuclear shape coexistence lies in the subtle interplay between two opposing trends, namely while shell or subshell closures stabilize spherical shapes, the residual interactions between valence protons and neutrons outside closed shells cause deformed configurations to become energetically favorable.  Studies of shape coexistence are important for understanding the microscopic origin of collectivity and the apparent disappearance of shell structures in exotic nuclei far from stability.

As summarized in Ref.~\cite{Heyde11}, the most pronounced spectroscopic fingerprints of coexistence between spherical and/or
shapes with quadrupole deformation are the diagonal $E2$ matrix elements, $B(E2)$ values, $E0$ transitions, isotope and isomer shifts, and two-nucleon separation energies. Among them, the diagonal $E2$ matrix elements and $B(E2)$ values are direct signatures of shape coexistence. In the case of even-even nuclei we are interested in the occurrence of significantly different shapes belonging to the ground state $0^+_1$ and the first excited $0^+$ state at relatively low energy. For $0^+$ states the diagonal $E2$ matrix elements vanish but, from a complete set of $E2$ matrix elements,  one can calculate model independent moments and higher order moments of the quadrupole operator -- the shape invariants.  Shape invariants were first introduced by Kumar \cite{Kumar72} and Cline \cite{Cline86} in the analysis of large sets of $E2$ matrix elements obtained in Coulomb excitation experiments.  In the geometrical model shape invariants can be related to the polar quadrupole deformation parameters $\beta$ and $\gamma$ or, to be more precise, to the effective values $\beta_{\rm eff}$ and $\gamma_{\rm eff}$ and their fluctuations. Quadrupole shape invariants have been extensively used to investigate shape coexistence in many regions of the nuclear chart \cite{Heyde11,Bree14,Suga03,Sreb06,Wrzo16,Zieli02,Ayang16,Garrett16}.

Studies of shape coexistence at low energy have evolved from isolated cases in deformed nuclei such as, for instance, prolate-oblate coexistence, to a generic phenomenon that occurs in nuclei over the entire mass table. It is, therefore, interesting to perform systematic analyses of characteristic signatures of coexisting shapes in different mass regions, particularly using a global self-consistent approach based on universal energy density functionals or effective interactions. In this work we present a calculation of quadrupole shape invariants for the two lowest $0^+$ states of even-even nuclei using a five-dimensional collective Hamiltonian (5DCH) with parameters determined by self-consistent relativistic mean-field calculations \cite{Nik09,Li09a}. This model goes beyond the simple mean-field approximation and takes into account correlations related to restoration of symmetries and fluctuations in collective coordinates. In a number of recent studies it has successfully been applied to analyses of low-energy collective states in various mass regions \cite{Nik09,Li09a,Meng16,Li09b,Li10,Li11,Li12,Lu15,Nik11,Pras12,Pras13}.


The lowest-order quadrupole invariants that characterize shape coexistence are defined by the following relations:
\begin{eqnarray}
\label{eq:q2}
q_2(0^+_i)&=&\langle0^+_i|Q^2|0^+_i\rangle=\sum\limits_j\langle0^+_i||Q||2^+_j\rangle\langle2^+_j||Q||0^+_i\rangle \\
\label{eq:q3}
q_3(0^+_i)&=&\sqrt{\frac{35}{2}}\langle0^+_i|Q^3|0^+_i\rangle
                 =\sqrt{\frac{7}{10}}\sum\limits_{jk}\langle0^+_i||Q||2^+_j\rangle\langle2^+_j||Q||2^+_k\rangle\langle2^+_k||Q||0^+_i\rangle \;.
\end{eqnarray}
These invariants can be related to the polar deformation parameters $\beta_{\rm eff}$ and $\gamma_{\rm eff}$:
\begin{eqnarray}
q_2(0^+_i)&=&\left(\frac{3ZeR^2}{4\pi}\right)^2\langle\beta^2\rangle\equiv \left(\frac{3ZeR^2}{4\pi}\right)^2\beta_{\rm eff}^2\\
\frac{q_3(0^+_i)}{q_2^{3/2}(0^+_i)}&=&\frac{\langle\beta^3\cos3\gamma\rangle}{\langle\beta^2\rangle^{3/2}}\equiv \cos3\gamma_{\rm eff}
\end{eqnarray}
where $R=r_0A^{1/3}$ and $r_0=1.2$ fm.

To compute quadrupole shape invariants, one has to evaluate the $E2$ matrix elements between states with angular momentum $0^+$ and $2^+$. To this end, we first carry out large-scale deformation-constrained self-consistent RMF+BCS calculation to generate mean-field single-nucleon wave functions in the entire $(\beta, \gamma)$ plane. The energy density functional PC-PK1 \cite{Zhao10} determines the effective interaction in the particle-hole channel, and a finite-range force that is separable in momentum space \cite{Tian09}, and adjusted to reproduce  the density dependence of the bell-shaped pairing gap in nuclear matter, is used in the particle-particle channel. The self-consistent Dirac equation for the single-particle wave functions is solved by expanding the solution in a set of eigenfunctions of a 3D harmonic oscillator potential in cartesian coordinates, including 12, 14, and 16 major shells for nuclei with $Z<20$, $20\leq Z<82$, and $Z\geq82$, respectively. The single-particle wave functions, occupation probabilities, and quasiparticle energies are used to calculate the mass parameters, moments of inertia, and collective potentials that determine the 5DCH, all of which are functions of the deformation parameters $\beta$ and $\gamma$. We note that the moments of inertia are calculated using the Inglis-Belyaev formula, and the mass parameters associated with the two quadrupole collective coordinates are determined in the perturbative cranking approximation.
The result of the diagonalization of the 5DCH \cite{Nik09,Li09a,Nik11} is the energy spectrum of collective
states and the corresponding eigenfunctions. The collective wave functions are used to calculate various observables, for instance, the quadrupole $E2$ reduced transition probabilities and spectroscopic quadrupole moments.

In the present analysis, for each $0^+$ state the sums in Eqs. (\ref{eq:q2}) and (\ref{eq:q3}) include the 30 lowest $2^+$ states, and this choice  ensures excellent convergence for the calculated quadrupole shape invariants. A systematic calculation of the invariants $q_2(0^+_1)$, $q_2(0^+_2)$, $q_3(0^+_1)$, $q_3(0^+_2)$, and the corresponding effective deformation parameters $\beta_{\rm eff}(0^+_1)$, $\beta_{\rm eff}(0^+_2)$, $\gamma_{\rm eff}(0^+_1)$, $\gamma_{\rm eff}(0^+_2)$ has been carried out for 621 even-even nuclei with $Z, N\geq10$, and for which the first $2^+$ state has been determined in experiment \cite{NNDC}. We consider the following criteria for shape coexistence: the difference between $\beta_{\rm eff}\cos3\gamma_{\rm eff}$ for the two lowest $0^+$ states is large, and the excitation energy of $0^+_2$ is low. To explore the role of triaxiality, we also consider the difference between $\beta_{\rm eff}$ for the two lowest $0^+$ states.

\bigskip
\begin{figure}[htb]
\includegraphics[scale=0.25]{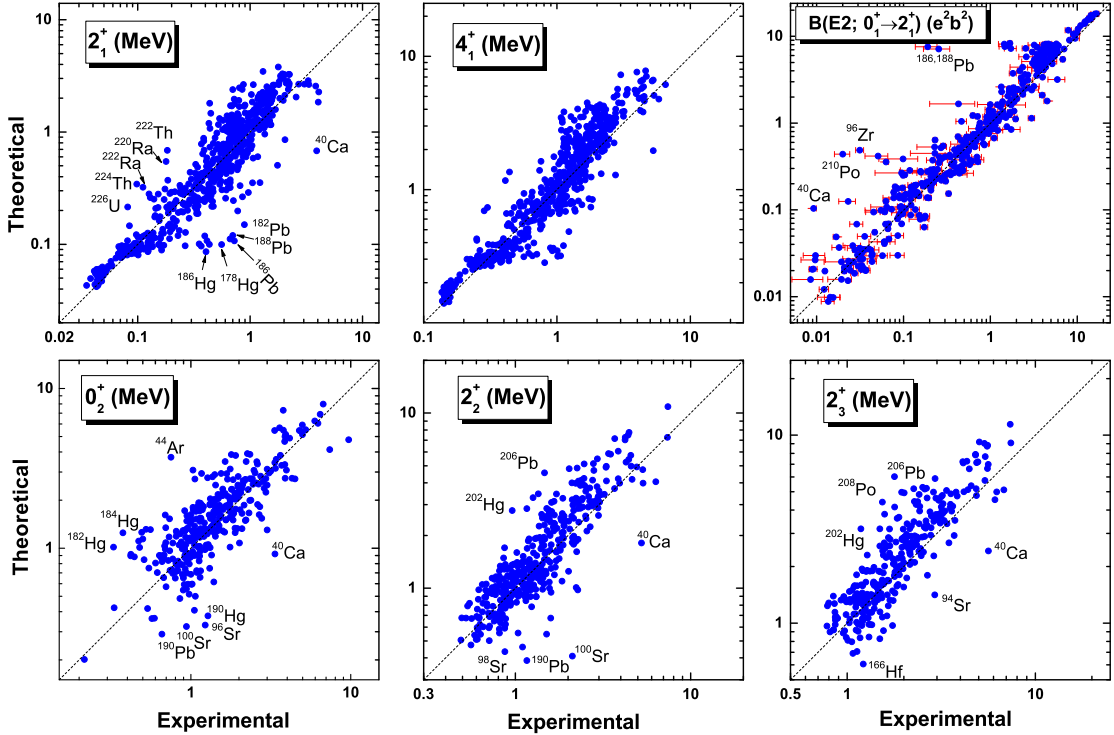}
\caption{(Color online) Theoretical excitation energies of the states $2^+_1$, $4^+_1$, $0^+_2$, $2^+_2$, $2^+_3$ , and
$B(E2; 0^+_1\to2^+_1)$ values (in units of $e^2b^2$) in even-even nuclei,
compared to the corresponding experimental values \cite{NNDC}.}
\label{fig:EBE2}
\end{figure}

Before discussing quadrupole shape invariants, in Fig.~\ref{fig:EBE2} we illustrate the quality of the 5DCH model calculation by comparing the theoretical  excitation energies of low-lying states $2^+_1$, $4^+_1$, $0^+_2$, $2^+_2$, $2^+_3$,  and the corresponding $B(E2; 0^+_1\to2^+_1)$ values,
to available data. The theoretical results are overall in good agreement with experiment, both for the excitation energies and $E2$ transitions,
especially considering that the excitation spectra have been calculated in the lowest order approximation with Inglis-Belyaev moments of inertia and
perturbative cranking mass parameters. For the $B(E2)$ values, in particular, the calculation not only reproduces the data in a wide interval of more than three orders of magnitude, but is also completely parameter free.  Namely, an important advantage of using structure models based on self-consistent mean-field single-particle solutions is the fact that observables, such as transition probabilities and spectroscopic quadrupole moments, are calculated in the full configuration space and there is no need for effective charges. This enables model calculations to reproduce empirical properties of
nuclei characterized by shape coexistence but also, more importantly, to make parameter-independent predictions in regions of exotic nuclei far from the valley of $\beta$-stability where few data are available. Exceptions are found in some transitional nuclei, and for nuclei with the number of protons $Z$ close to the magic numbers. One of the principal reasons is that the present 5DCH model includes only quadrupole collective degrees of freedom and,
therefore, does not explicitly take into account two-quasiparticle configurations and, for instance, octupole deformations.
We note that the overall quality of the results shown in Fig.~\ref{fig:EBE2} is comparable to those obtained using
the Gogny D1S effective interaction in Refs.~\cite{Del10,D1S}.

\begin{figure}[htb]
\includegraphics[scale=0.5]{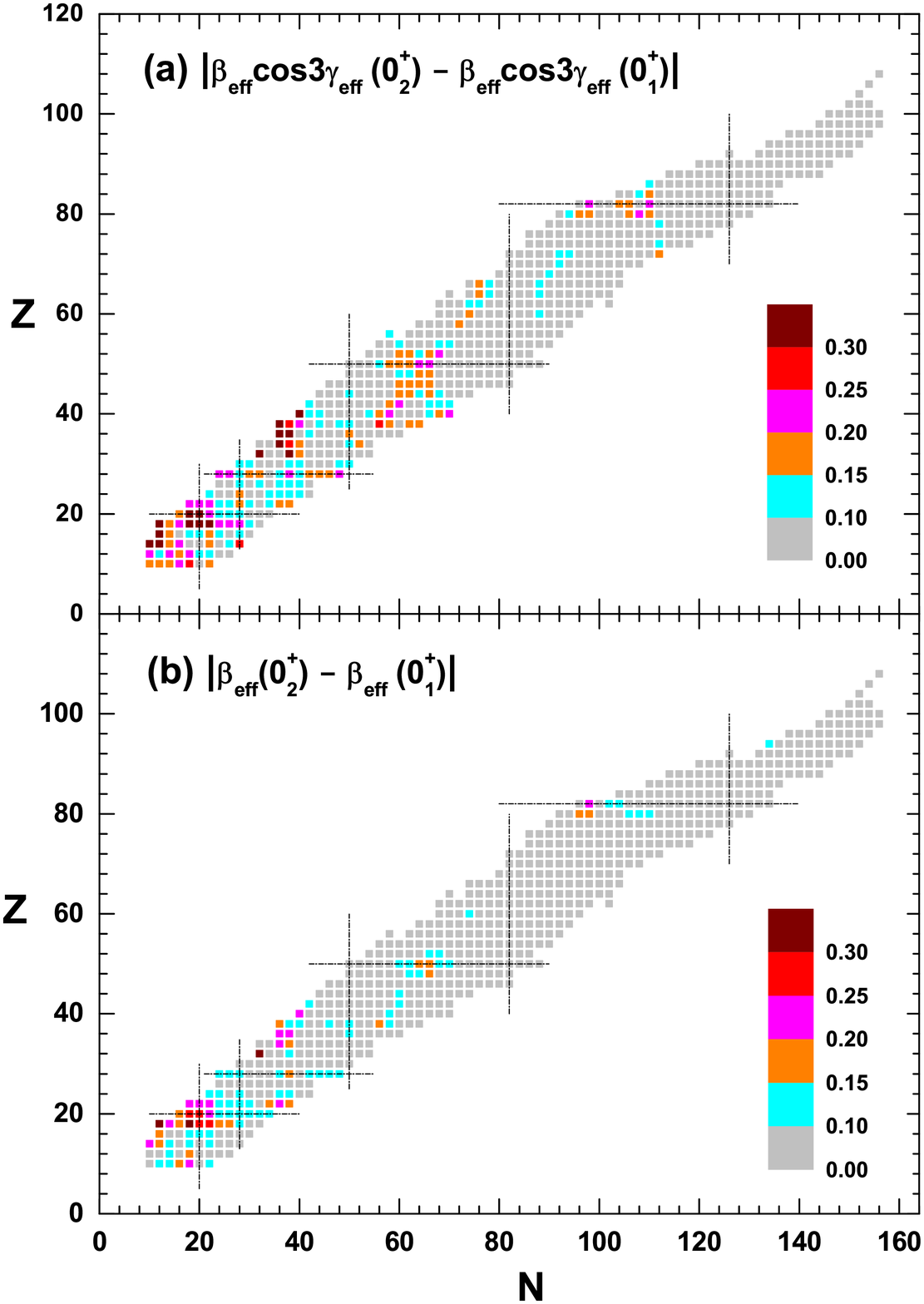}
\caption{\label{fig:bg21}(Color online) Absolute differences between the calculated $\beta_{\rm eff}\cos3\gamma_{\rm eff}$ (a) and $\beta_{\rm eff}$ (b)
values for the two lowest $0^+$ states of 621 even-even nuclei.}
\end{figure}
\begin{figure}[htb]
\includegraphics[scale=0.5]{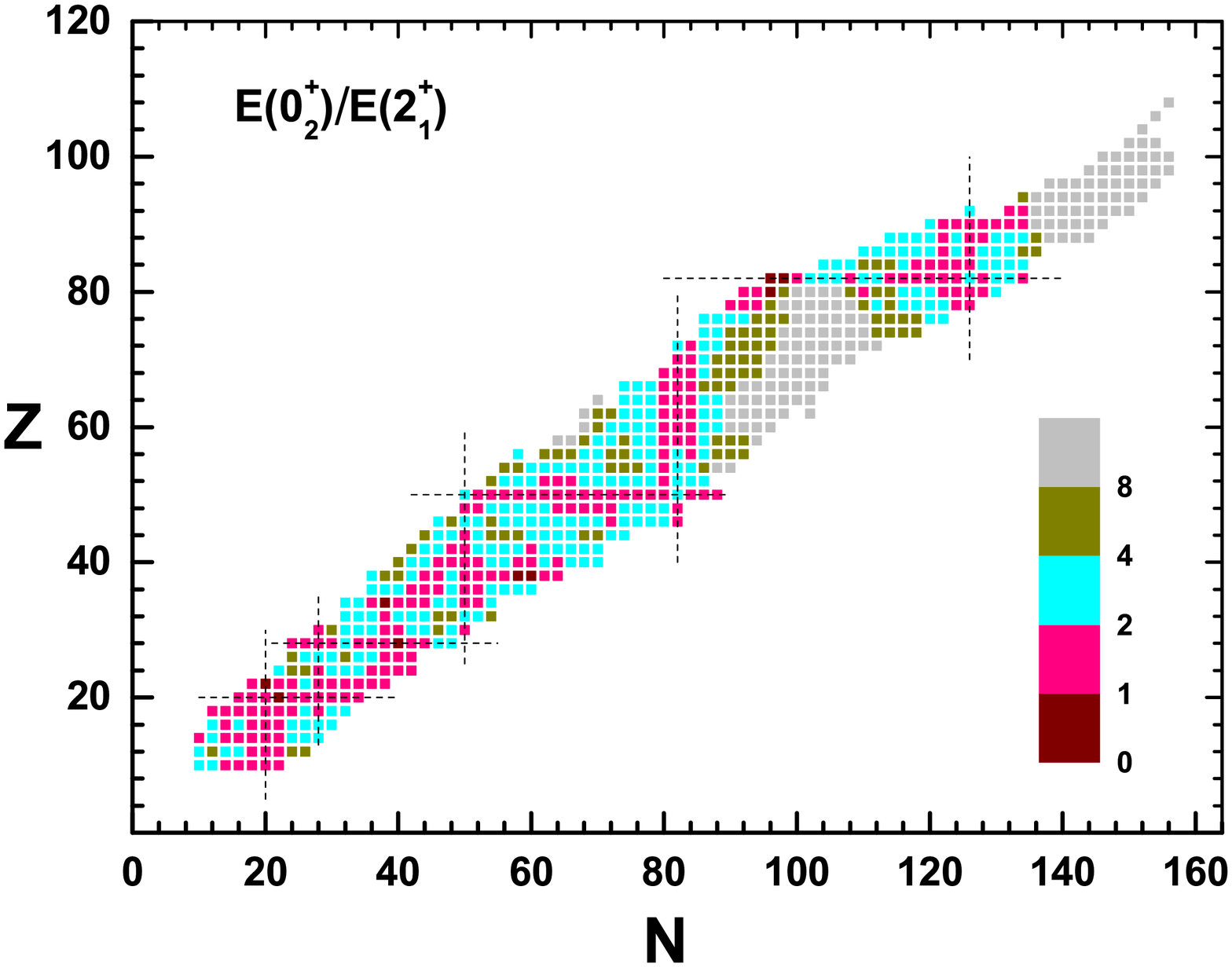}
\caption{\label{fig:E02}(Color online) The calculated ratios between the excitation energies of the states $0^+_2$ and $2^+_1$ for 621 even-even nuclei.}
\end{figure}

Figure \ref{fig:bg21} displays the calculated absolute differences of $\beta_{\rm eff}\cos3\gamma_{\rm eff}$ (a), and $\beta_{\rm eff}$ (b)
between the two lowest $0^+$ states for 621 even-even nuclei. We also map the ratios between the excitation energies of the states  $0^+_2$ and $2^+_1$ in Fig.~\ref{fig:E02}. Shape coexistence may be expected to occur in nuclei in which absolute values of these differences are relatively
large, for instance $>0.1$ and, simultaneously, the first excited $0^+_2$ states are low in energy when compared to the excitation energy of
$2^+_1$. Overall, the distribution of possible shape-coexisting nuclei shown in Fig.~\ref{fig:bg21} (a) and (b), and Fig.~\ref{fig:E02},
is consistent with the main regions of shape coexistence
summarized by Heyde and Wood in Fig.~8 of their review paper \cite{Heyde11}. Here the aim is not to compute properties of individual
shape-coexisting nuclei, but to indicate mass regions that display structure properties associated with the phenomenon of shape coexistence.
Based on the results obtained in the present analysis, we discuss the occurrence of shape coexistence in different regions of the table of nuclides:
\begin{enumerate}
\item[(i)]  Nuclei in the vicinity of $Z\sim50$ and $Z\sim82$. Coexistence of low-lying spherical and intruder deformed shapes has been
extensively studied and demonstrated by numerous experiments
in Sn, Cd, Te, Pb, Hg, and Po isotopes \cite{Heyde11,Garrett16,Wrzo16,Bree14,Gaffney14,Gaffney15,Else11,Chen16,Julin16}. Shape
coexistence in these regions can also be related to triaxiality, but the number of possible triaxial shape-coexisting nuclei is not large.
\item[(ii)] $Z\sim64, N\sim90$ nuclei. The primary interest in this region is the rapid onset of deformation in the transition from $N=86$ to $N=92$ \cite{Cejnar10,Li09a,Li09b}. The issue of shape coexistence here is somewhat subtle because there are no obvious differences in band energy spacing or $B(E2)$ values. Moreover, the $0^+_2$ states are found at relatively high energies because of strong mixing between the two lowest $K=0$ bands.
\item[(iii)] $Z\sim64, N\sim76$ nuclei. Medium-deformed triaxial ground states coexisting with highly-deformed prolate excited state are predicted in this region. Furthermore, it is found that triaxial ground states originate from the interaction between proton multi-particle and neutron multi-hole states, and the prolate excited states are built on a deformed neutron shell gap with $\beta\sim0.4$. This result is consistent with that obtained using the Gongy D1S effective interaction \cite{Del10,D1S}. New measurements of spectroscopic properties are suggested in this mass region, especially for the nuclei  $^{134}$Nd, $^{136,138}$Sm, $^{140,142}$Gd, and $^{142,144}$Dy.
\item[(iv)] $Z\sim40, N\sim60$ nuclei. The structure of nuclei in this mass region is characterized by a sudden onset of deformation in the transition from $N=58$ to $N=60$, as demonstrated by the dramatic change in the isotope shifts $\delta\langle r^2\rangle$ and two-neutron separation energies $S_{2n}$. These changes occur because of the crossing between coexisting structures, that is, highly-deformed prolate and spherical configurations \cite{Heyde11}. Numerous  measurements of spectroscopic quadrupole moments, $B(E2)$ values, $E0$ transitions, two-nucleon and $\alpha$-cluster transfer data, have revealed the onset of shape coexistence in Sr, Zr, and Mo isotopes \cite{Heyde11,Kremer16, Clem16,Clem16b,Clem16c,Chak13,Thomas13,Thomas16,Gorgen16,Park16}, while static and dynamic quadrupole moments data show that shape coexistence still persists in Ru and Pd isotopes \cite{Heyde11,Sreb06,Smith12,Wang00,Sven96}. Our calculation also indicates that heavier Ru and Pd isotopes exhibit shape coexisting structures, with triaxiality playing an important role \cite{Sven96,Doherty17}.
\item[(v)] $Z\sim40, N\sim70$ nuclei. These nuclei are very neutron-rich and only limited spectroscopic information is
 available. Recently the first measurement of low-lying states in the neutron-rich $^{110}$Zr and $^{112}$Mo was performed 
 via in-beam $\gamma$-ray spectroscopy. Low-lying  $2^+_1$ states observed at excitation energies 185(11) and 235(7) keV, 
 respectively, as well as $R_{42}$ values $\sim3$, indicate that both nuclei are well deformed \cite{Paul17}. The present study has also shown that $^{110}$Zr does not exhibit a stabilizing shell effect corresponding to the harmonic oscillator magic numbers $Z=40$ and $N=70$, thus pointing to possible shape coexistence in this mass region. We note that the present self-consistent mean-field calculation predicts a spherical global minimum in $^{110}$Zr, similar to other recent mean-field results \cite{Zhao17}. The deformed ground state is obtained by taking into account dynamical, beyond mean-field, correlations.
\item[(vi)] $Z\sim34, N\sim40$ nuclei. The manifestation of shape coexistence in this region can be attributed to three shell gaps in the Nilsson level diagram: a weakly-deformed shell gap at $Z=34$, a spherical subshell closure at $N=40$, and a highly-deformed prolate shell gap at $N=38$ \cite{Fu13,Wang15}. Detailed discussions of spectroscopic properties can be found in Refs. \cite{Heyde11,McCu13,Corsi13,Ayang16,Gorgen16}, and references therein. We note that, according to Fig. \ref{fig:bg21} (a), the $Z\sim40$ and $N\sim40$ nuclei also display shape-coexisting structures, similar to predictions in Refs.~\cite{Toma11,Zheng14}. However, the $0^+_2$ excitation energies in these nuclei are rather high (c.f. Fig. \ref{fig:E02})  and, therefore, additional measurements are necessary to clarify these structures.
\item[(vii)] $Z\sim28$ nuclei. Shape coexistence is observed in nuclei with $N\sim28, 40, 50$ \cite{Rudo99,Dijon12,Suchyta14,Prokop15,Leoni17,Gott16,Yang16}, which is consistent with our predictions.
\item[(viii)] $(N, Z)\sim(20, 12), (28, 14)$, and $(40, 24)$ nuclei. The so-called ``islands of inversion'' have attracted considerable interest in the last two decades. These studies have been summarized in Refs. \cite{Heyde11,Gade16}. The term island of inversion refers to the fact that 2p-2h states are located below 0p-0h closed-shell states. This implies inversions of states, and results in phenomena that basically do not differ from well known structures characterized by shape coexistence. The results obtained in the present study are generally consistent with measurement except for the $N=20$ isotones, for which the $N=20$ shell closure calculated with the functional PC-PK1 is simply too strong.
\item[(ix)] $N\approx Z$ light nuclei. Our predictions are consistent with the nuclei listed in Table III of Ref. \cite{Heyde11}, especially for the $^{40}$Ca region.
\item[(x)] $N, Z<20$ nuclei. The occurrence of shape coexistence is predicted in many nuclei in this region due to a rather large effect of quadrupole deformation \cite{Del10,Zhang14}.
\end{enumerate}

In summary, we have performed a systematic calculation of quadrupole shape invariants for the two lowest $0^+$ states of 621 even-even nuclei.
Excitation spectra and E2 transition matrix elements have been computed using the five-dimensional collective Hamiltonian model based on
the relativistic energy density functional PC-PK1, and a finite-range pairing interaction. The model accurately reproduces available data on
excitation energies of the low-lying states $2^+_1$, $4^+_1$, $0^+_2$, $2^+_2$, $2^+_3$, and $B(E2; 0^+_1\to2^+_1)$ values, over the entire chart
of nuclides. The resulting quadrupole shape invariants $q_2$ and $q_3$ for the states $0^+_1$ and $0^+_2$ can be related to the
corresponding effective polar deformation parameters $\beta_{\rm eff}$ and $\gamma_{\rm eff}$. A systematic comparison of shape invariants for the
lowest $0^+$ states indicates regions of possible shape coexistence. Coexistence of different geometric shapes at low energies has emerged as a
universal structure phenomenon that occurs in different mass regions over the entire chart of nuclides.  A global theoretical approach such as the one based on energy density functionals is essential for accurate predictions in regions far from stability
where few data are available. In this work signatures of
shape coexistence have been analyzed and compared with previous theoretical studies and available data. It has been shown that, when based on
a universal and consistent microscopic framework of nuclear density functionals, shape invariants provide distinct indicators and reliable predictions for the occurrence of low-energy coexisting shapes.

\begin{acknowledgements}
This work was supported in part by the NSFC under Grant Nos. 11475140 and 11575148,
the Croatian Science Foundation -- project ``Structure and Dynamics
of Exotic Femtosystems" (IP-2014-09-9159), the QuantiXLie Centre of Excellence,
and the Chinese-Croatian project ``Microscopic Energy Density Functionals Theory for Nuclear Fission''.
\end{acknowledgements}


\end{document}